# Shift in Nomenclature not Thesis: Innate Immune Memory, Pathogen Dose and Opsonins

Donald R. Forsdyke[1*]

[1]Department of Biomedical and Molecular Sciences, Queen's University, Kingston, Ontario, Canada K7L3N6

[*]Correspondence: forsdyke@queensu.ca

## Early-Onset Responses to Infection

Recent writings on "innate immune memory" or "trained immunity," and on "hormetic responses" [1-3], herald reinvigoration of an important research area with early 20[th] century roots. However, it is questionable that the thesis of a "major importance of the sensed dose of pathogen for the development of innate immune system-mediated responses," should be labelled a "paradigm shift" [1]. The works of Almroth Wright (1861-1947) should be taken into account. In his time it was known that there were at least three early-onset, relatively non-specific, responses to infections – fever (pyrexia), enhanced aggregation of red blood cells (rouleau formation), and an increase in white blood cells (leukocytosis). Pyrexia was later shown to vary with pathogen dosage, to have a memory component, and to differ in kinetics from specific antibody responses (reviewed in [4]). Rouleaux formation was later shown to reflect changes in the aggregating power of plasma proteins (reviewed in [5]).

## Opsonins as an Early Adaptive Response

Whether these responses to pathogens were positively adaptive was unknown to Wright when, around 1900, he began studies of a fourth early-onset response in humans. This was the increase in the plasma "opsonins" that enhanced ingestion of bacteria by certain white blood cells – the polymorphonuclear phagocytes. In that the bacteria were destroyed, the adaptive value of their "opsonization" was not in doubt. Wright distinguished his relatively non-specific opsonins from the highly pathogen-specific antibodies that appeared later in infections. Although the kinetics were different, as with "acquired" immune antibody responses, "innate" immune opsonin responses also had an acquired (i.e. memory) component (Fig. 1). Furthermore, just as antibody responses were influenced by pathogen dose, opsonin responses were similarly susceptible (reviewed in [6]).

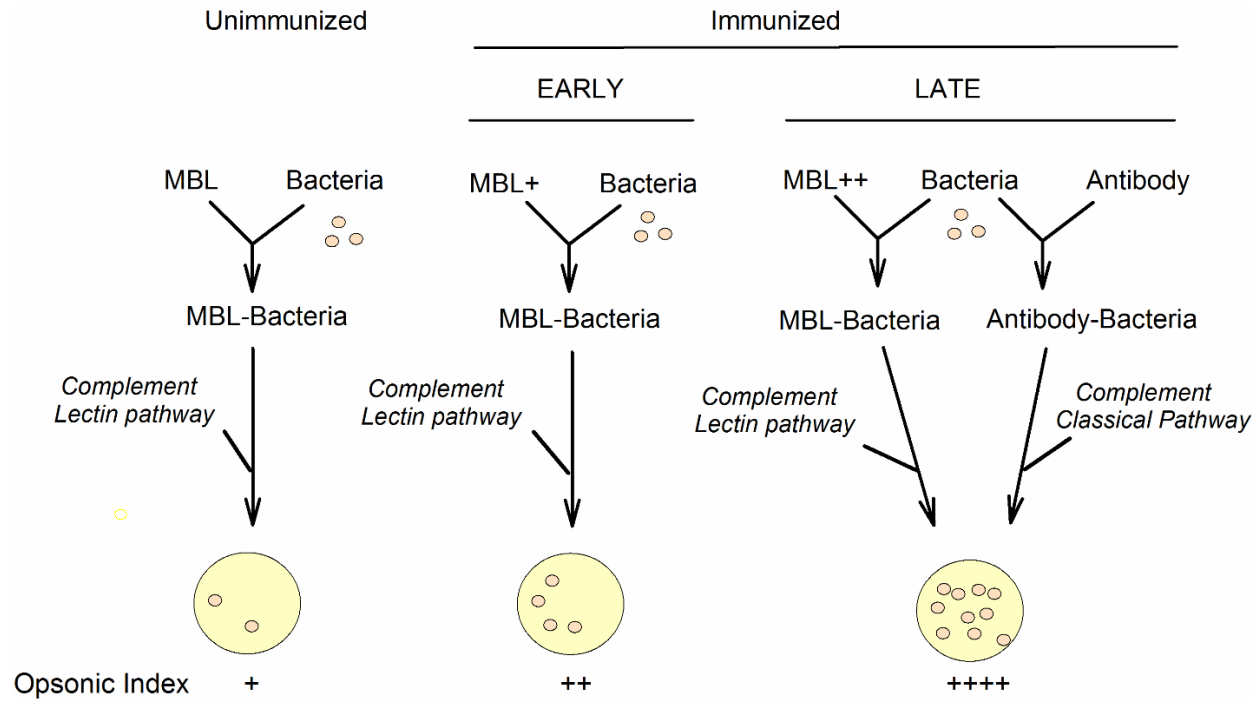

**Figure 1**. A summary of Wright's view of opsonisation. It is assumed that his heat-labile activity is complement and his heat-stable "bacteriotropic" substances that activate the lectin complement pathway are mannose-binding lectins (MBLs) and/or related non-antibody molecules (e.g. ficolins). Orange circles represent bacteria. Yellow circles represent the phagocytes that ingest bacteria when they are appropriately opsonized. Reproduced from [6] with permission.

**Father of Innate Immunity?**

Although researchers may engage in a "thorough … search of the literature" [2], many accounts do not alert them to Wright`s studies (e.g. see [7]). For some in this field it suffices to be guided by Janeway's famous 1989 paper on innate immunity [8], and to accept his view that "innate immunity has usually been treated as a minor curiosity." Backing this, Janeway explained "how we arrived where we are," called for "a rediscovery of microbiology by immunologists," and postulated that a complement-based defensive "effector mechanism existed prior to the development of specific antigen receptors" (i.e. before antibody-dependent complement-activating mechanisms evolved). This is vintage Wright. The elegant new work [1-3] now adds interesting twists to Janeway's thesis, including the addition of a memory component. However, while some regard Janeway as "the father of innate immunity" [9], the validity of so recent a paternity is debateable [10].

**Shifts in Nomenclature**

Terms such as "trained immunity" and "hormesis" [1], or "pattern recognition receptors" (PRRs) and "pathogen associated molecular patterns" (PAMPs) [8]), may be novel, but the underlying thesis is not. The Google N-gram viewer indicates that the terms "hormesis" and "hormetic" surfaced in the 1940s, but did not really take off until the mid-1980s, when PRRs and PAMPS surfaced. Although these words were not employed, studies initiated in the 1960s with mannose-binding lectins (MBLs) – now recognized as major components of innate immune systems [11] – produced "hormetic" results (i.e. stimulation at low concentrations and complement-dependent inhibition at high concentrations; reviewed in [6]). The heat-stable components of Wright's opsonins have since been characterized at the molecular level (MBLs, ficolins, etc.) and these components have their own "lectin pathway" to his heat-labile activity – complement [12, 13].

The opsonization shown in Figure 1 primarily affects neutrophil polymorphonuclear phagocytes. While correctly observing that: "Three main populations function as cellular effectors of innate immunity: the neutrophils, the monocytes/macrophages and the NK cells," it has been concluded [14]: "Neutrophils are cells in a state of final differentiation and a short life span" and thus "inherently their capacity to participate in long term immunological memory is very limited." However, as implied by Wright's term "opsonization" [6], he had found neutrophils to be passive receivers. They were armed by the plasma opsonins that increased following an infection. Wave upon wave, neutrophils would perish, by not before they could respond to the dictates of these soluble PRRs. We should not now refer to this as a "previously unrecognized property of human immune responses" [15].

513. Andrade, F.A. *et al.* (2017) Serine proteases in the lectin pathway of the complement system. In *Proteases in Physiology and Pathology* (Chakraborti, S. and Dhalla, N.S., eds.), pp 397–420, Springer Nature

14. Netea, M.G. (2013) Training innate immunity: the changing concept of immunological memory in innate host defence. *Eur. J. Clin. Invest*. 43, 881–884

15. Netea, M.G. *et al*. (2016) Trained immunity: A program of innate immune memory in health and disease. *Science* 352, 427